\documentclass{aa501}
\usepackage{graphicx}
\begin{document}
   \title{On the double-mode RR Lyrae variables of the Sculptor 
          dwarf galaxy}

   \author{G. Kov\'acs
          \inst{}}

   \offprints{G. Kov\'acs}

   \institute{Konkoly Observatory, P.O. Box 67, H-1525, Budapest, Hungary\\
              \email{kovacs@konkoly.hu}}

   \date{Received 22 February 2001 / Accepted 12 June 2001 }

   \titlerunning {Double-mode RR Lyrae stars of Sculptor}

   \abstract{
Frequency analysis of more than 300 stars of the {\sc ogle} data base 
on Sculptor galaxy has led to the discovery of 18 double-mode RR~Lyrae 
(RRd) variables. This yields a 20\% incidence rate for double-mode pulsation 
among the variables previously classified as first overtone RR~Lyrae 
stars in this galaxy. Most of the RRd stars cover the period range 
of $0\fd47<P_0<0\fd49$ but there are two stars with longer periods of 
$\approx 0\fd54$. All variables fit well in the pattern of the 
$P_0\rightarrow P_1/P_0$ diagram, spanned by the RRd stars of the 
Galactic globular clusters and those of the Large Magellanic Cloud (LMC). 
It follows from our previous investigations that the luminosities and 
masses of the RRd stars in Galactic globular clusters and in the LMC 
are almost independent of metallicity. By assuming that the Sculptor 
RRd variables also obey this rule, with the aid of the pulsation 
equations we estimate their metallicities. For most of the stars we 
get [Fe/H]$\approx-1.5$, which is the same value as that obtained 
from a semi-empirical method for the average metallicity of the 
fundamental mode (RRab) stars. Two RRd stars have considerably lower 
metallicities, but even those are within the range corresponding to 
the RRab stars. The narrower metallicity range of the RRd stars is in 
agreement with their observed luminosity range, which is about three 
times smaller than that of the RRab stars.       
   }

   \maketitle
%

%
%

\section{Introduction}
The variable star content of the Sculptor dwarf spheroidal galaxy has 
been investigated recently as a `target of opportunity' by the {\sc ogle} 
project, devoted primarily to searches for microlensing events 
(Kaluzny et al. 1995). They found 326 stars suspected of variability, 
from which 134 variables were classified as fundamental mode (RRab) and 
89 stars as first overtone (RRc) RR~Lyrae variables. They also noted the discovery of one double-mode (RRd) star. The purpose of the present paper 
is to conduct a more thorough search for RRd stars in the {\sc ogle} data 
base and study their physical properties. 

In a recent series of papers we have shown that double-mode variables are 
very useful objects in giving fairly accurate estimates on the distance 
moduli of benchmark systems, such as the Magellanic Clouds (Kov\'acs \& 
Walker 1999; Kov\'acs 2000a, b; hereafter KW99, K00a and K00b, 
respectively). A complete parameter estimation of an 
RRd star requires not only the knowledge of the periods but also that of 
two additional parameters, such as the color index and metal abundance. 
Unfortunately, in the case of the present data set we have light curves 
available only in the $V$ band. Therefore, it is necessary also to use  
some other information, in order to estimate the allowed parameter 
regimes for the Sculptor variables. 

This additional information comes from our earlier studies on the 
RRd populations in various Galactic globular clusters. In those works,  
multicolor photometry was available, therefore, we could obtain 
mass and luminosity estimates. Here we examine these estimates more 
closely. We will see that they exhibit marginal (if any) metallicity 
dependence. By adopting these masses and luminosities in the present 
study, and applying the pulsation equations, we give estimates on the metallicities of the Sculptor RRd variables.

%
%
 
\section{Frequency analysis}
The method of analysis is very similar to the one used by Alcock et al. 
(2000, hereafter A00). The main difference is in the calculation 
of the frequency spectra. Here we use a single component least squares 
(LS) criterion (e.g., Barning 1963) rather than the standard Fourier 
transform method (e.g., Deeming 1975). We opted for the LS approach 
because in the case of the present data set with high noise and strong 
aliasing, this method yields a better estimate for the peak frequency 
position. The difference between the two methods becomes unimportant, 
e.g., in the case of RRd variable searches performed on the type of 
data sets analyzed by A00.  
%
%
%
   \begin{figure}[t]
   \centering
   \includegraphics[width=90mm]{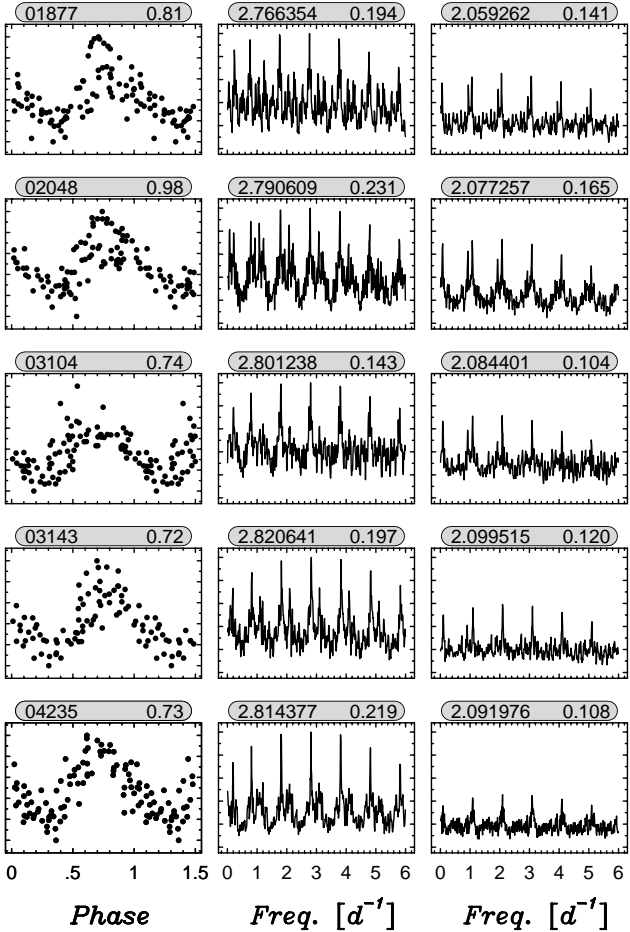}
      \caption{Sample spectra and folded light curves for some of the 
most securely identified RRd variables in Sculptor. Each row displays 
the result of the analysis of the variable identified by the very 
first item of data in the respective header. {\it First box:} 
V light curve folded with the period of the highest amplitude. 
{\it Second box:} LS amplitude spectrum of the original data.  
{\it Third box:} LS amplitude spectrum of the data prewhitened by the 
highest amplitude component shown in the second box. The headers show 
(from left to right): {\sc ogle} identifier, total range of the 
light variation, frequency and the corresponding amplitude for the 
original and prewhitened data. Amplitudes are given in [mag] }
         \label{}
   \end{figure}
%
%
%
   \begin{figure}[t]
   \centering
   \includegraphics[width=90mm]{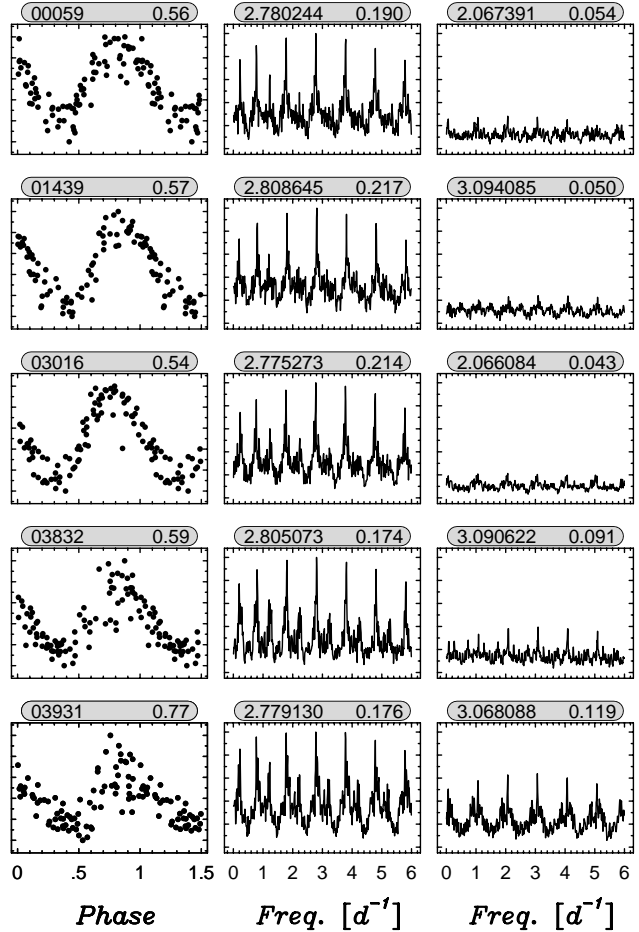}
      \caption{Sample spectra and folded light curves of some of the 
less securely identified RRd variables in Sculptor. Notation is the 
same as in Fig. 1 }
         \label{}
   \end{figure}

Secondary frequency components are searched by prewhitening 
the signal with the highest amplitude component. The full data set, 
containing 326 variables, is analyzed. Each time series spans some 425 
days with $\approx 80$ data points in Johnson $V$ color. The data 
are concentrated in two observing seasons, separated by one year. The 
first season covers only three weeks and contains about 15\% of the total 
number of observations. The second season covers a longer time interval, 
spanning of about 2.5 months. There is also a strong 1-day periodicity 
in the sampling, which affects the frequency spectra rather severely. 
However, in the present analysis, this problem is less disturbing, 
because nonphysical frequencies can be excluded relatively easily, due 
to the narrow period ratio range of the RRd stars ($0.741<P_1/P_0<0.748$, 
if we consider {\it all} presently known RRd variables --- see Popielski 
et al. 2000).

By visually checking all frequency spectra, 18 variables are selected 
as RRd candidates, based on the patterns of the prewhitened spectra. 
For illustration, we display some of the most secure identifications 
in Fig.~1. It is seen that the daily aliasing is indeed strong. 
Even so, the LS method is capable of identifying the correct components.  
(By using a standard discrete Fourier transform method, we find that 
three out of five stars are misidentified.) In some other cases the 
frequency search leads to less reliable results. Fig.~2 displays some 
examples of this. 

The lower two panels exhibit the ambiguities from the daily aliasing. 
In these cases physical arguments are applied to choose among the 
various possibilities. For example, in the case of variable \#03832, 
the ratio of the formally obtained frequencies is 0.908, which excludes 
the possibility of both components identified as radial modes. However, 
the frequency difference is 0.286$d^{-1}$, which is too large for this 
star to be one of the newly discovered variables of RRc stars with 
closely spaced frequencies (A00; Olech 1999). On the other hand, we get 
a very simple explanation for the observed pattern if we assume that 
the true secondary component is at $(3.090622-1.0)d^{-1}$. This assumption 
leads to a period ratio of 0.7443, which is in the right range for an 
RRd variable. 

%
%
%
\setcounter{table}{0}
\begin{table}[h]
\caption[ ]{Periods, amplitudes and intensity-averaged magnitudes 
of the RRd stars in the Sculptor 
dwarf galaxy}
\begin{flushleft}
\tiny
\begin{tabular}{rcccccc}
\hline
Name   & $P_0$ & $P_1$ & 
$P_1/P_0$ & 
$A_1$ & 
$A_0/A_1$ & 
$\langle V \rangle$   \\
\hline
 $^\star$00059 
        &  0.48370  &  0.35968  &  0.7436  &  0.190  &  0.28 & 20.16 \\
 01168  &  0.54153  &  0.40357  &  0.7452  &  0.187  &  0.70 & 19.99 \\
 $^\star$01439 
        &  0.47809  &  0.35604  &  0.7447  &  0.217  &  0.22 & 20.16 \\
 01877  &  0.48561  &  0.36149  &  0.7444  &  0.194  &  0.73 & 20.17 \\
 02048  &  0.48140  &  0.35835  &  0.7444  &  0.231  &  0.71 & 20.17 \\
 $^\star$03016 
        &  0.48401  &  0.36033  &  0.7445  &  0.214  &  0.20 & 20.13 \\
 03044  &  0.47546  &  0.35427  &  0.7451  &  0.168  &  0.58 & 20.12 \\
 03104  &  0.47975  &  0.35699  &  0.7441  &  0.143  &  0.73 & 20.17 \\
 03143  &  0.47630  &  0.35453  &  0.7443  &  0.197  &  0.61 & 20.20 \\
 03832  &  0.47894  &  0.35650  &  0.7443  &  0.174  &  0.56 & 20.19 \\
 03931  &  0.48356  &  0.35983  &  0.7441  &  0.176  &  0.68 & 20.20 \\
 03941  &  0.47915  &  0.35661  &  0.7442  &  0.188  &  0.47 & 20.15 \\
 04235  &  0.47802  &  0.35532  &  0.7433  &  0.219  &  0.50 & 20.22 \\
 04353  &  0.48298  &  0.35983  &  0.7450  &  0.181  &  0.51 & 20.15 \\
 04824  &  0.48658  &  0.36222  &  0.7444  &  0.223  &  0.46 & 20.09 \\
 05354  &  0.54074  &  0.40322  &  0.7457  &  0.177  &  0.56 & 20.10 \\
 05730  &  0.47254  &  0.35189  &  0.7447  &  0.180  &  0.37 & 20.20 \\
 05845  &  0.48537  &  0.36135  &  0.7445  &  0.228  &  0.50 & 20.15 \\
\hline
\end{tabular}
\end{flushleft}
{\footnotesize
\underline {Note:}
Variables labeled by $\star$ are marginal cases for RRd classification}
\end{table}
%
%
%
%
%
   \begin{figure}
   \centering
   \includegraphics[width=90mm]{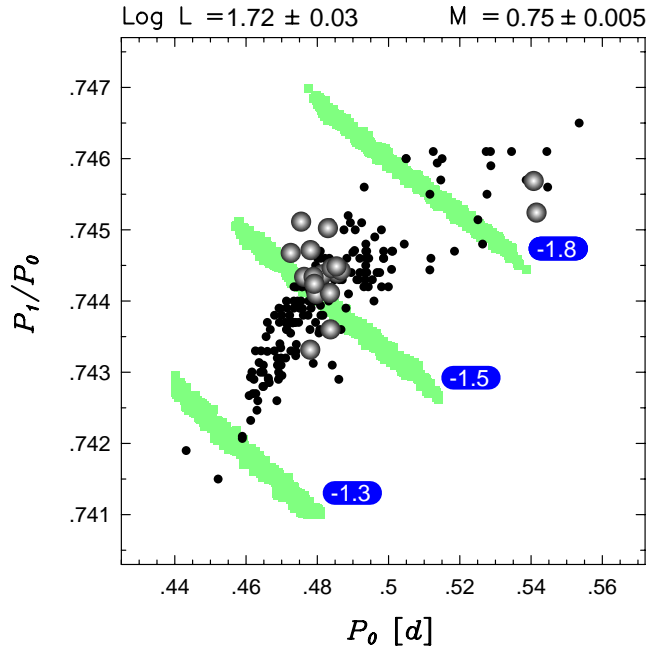}
      \caption{Position of the Sculptor RRd variables (spheres) on the 
$P_0\rightarrow P_1/P_0$ diagram. The LMC RRd variables 
are plotted by dots. Shaded areas show the position of the theoretical 
models computed by the parameters given in the header and [Fe/H] values 
displayed at the corresponding ridges. In order to place the middle of 
the ridges on the observed points, effective temperatures are fixed at 
6900, 6800 and 6700~K for the [Fe/H]=$-1.3$, $-1.5$ and $-1.8$ models, 
respectively. The {\sc opal} opacities given by Iglesias \& Rogers (1996) 
are used with $X=0.76$}
         \label{}
   \end{figure}

The other three stars in Fig.~2 have rather low-amplitude secondary 
components. This makes it somewhat difficult to identify them as RRd 
variables. The primary reason why they have been selected as probable 
RRd stars is that they have `correct' period ratios and their fundamental 
periods are sufficiently far from $0\fd5$. Other RRd candidates (i.e., 
\#00038, \#03834 and \#05015) with similarly low fundamental amplitudes 
have been rejected because they did not satisfy this criterion. When  
considering these questionable cases, we have to keep in mind that the 
analysis of all these variables is not yet final, because the presently 
available data are rather limited. Nevertheless, we do not think that 
considerable changes will take place in the status of most of the Sculptor 
RRd stars, even when more extended data become available.   

The basic observed properties of the RRd variables identified in this 
search are listed in Table~1 (periods are in [d], the $A_1$ 
amplitudes of the first overtone components and the average magnitudes 
in [mag]). We note that the periods and amplitudes displayed in this 
table have been computed through separate LS fits, which contained the 
given component and its first harmonics. We see that there are only 
two stars with periods of $\approx 0\fd54$; the other variables occupy 
a narrow range of $0\fd47<P_0<0\fd49$. We recall that variable \#01168 
was already discovered by Kaluzny et al. (1995). 

From the periods alone, we may suspect that the physical properties of 
the Sculptor RRd variables cover similar ranges to those of the globular 
cluster IC4499. Indeed, estimated from the RRab stars, the two systems 
have almost the same average metallicity of $-1.5$ (see Kov\'acs \& Walker 
2001\footnote{Please note that the reddened relative distance modulus 
$\langle\tilde d_V\rangle$ of M4 in Table~9 of this paper is misprinted. 
The correct value is $-4.43$.}. Furthermore, a display of the Sculptor 
RRd variables on the $P_0\rightarrow P_1/P_0$ diagram (often referred 
to as the `Petersen diagram' --- see Petersen 1973), shows that they 
occupy a region which corresponds to this metallicity (see Fig.~3). 
The main factor of degeneracy in this diagram comes from the similar 
effects of the metallicity and mass on the period ratio. Therefore, 
as we will see in the next section, if no other information is available, 
the periods can be fitted with any metallicity between $-1.3$ and $-2.1$, 
if we allow masses between $0.65$ and $0.85M_{\sun}$. It is clear that 
for a more reliable assessment of the physical parameters of the Sculptor 
RRd variables, we would need direct metallicity measurements and 
additional color information in order to fix the temperature and 
chemical composition, and thereby to provide an (almost) complete 
parameter input for the solution of the pulsation equations (cf., K00b). 
Since none of these data are available at this moment, in the next 
section we follow another idea, which adopts the almost metallicity-
independent luminosity and mass values obtained from our former studies 
on RRd stars.

%
%
 
\section{Computation of the metallicity}
In K00b we derived the distance modulus of the LMC by directly using  
its RRd population. The compatibility between this distance modulus 
and the ones obtained from the RRd stars of Galactic globular clusters 
and Small Magellanic Cloud Cepheids (see K00a) was also shown. Because 
of the consistency of all these results, we think that the physical 
parameters to be derived are also reliable and can be extended to the 
Sculptor RRd variables. 

Metallicities for the Galactic globular clusters were adopted from KW99, 
where it was demonstrated that the condition of fixed relative distance 
moduli (obtained from the RRab stars with the method of Kov\'acs \& 
Jurcsik 1997) yields metallicities very close to the generally accepted 
overall cluster values. In the case of low metallicity clusters this 
method does not result in a strong constraint, because of the insensitivity 
of the period ratios to metallicity at such low abundances. Therefore, 
for M15 we took [Fe/H]=$-2.3$, which may be regarded as a value close 
to those often quoted in the literature. We will see that a somewhat 
higher value would be perhaps more appropriate for this cluster. For LMC 
we used the very recent spectroscopic data of Clementini et al. (2000) 
to set its average metallicity at [Fe/H]=$-1.5$.    
      
Table~2 lists the average luminosity and mass values obtained in the 
way given in KW99 and K00b. We mention that {\sc opal} opacities of 
Iglesias \& Rogers (1996) are applied with hydrogen abundance 
$X=0.76$ and a solar-type heavy element distribution (i.e., without 
oxygen enhancement) in the overall heavy element content. The zero point 
of the color--temperature transformation is tied to the temperature 
scale defined by the current results of the infrared flux method 
(Blackwell \& Lynas-Gray 1994). For completeness, the metallicities and 
effective temperatures are also listed in the table. We note that this 
latter quantity is basically independent of the pulsation models and 
is determined only by the adopted color--temperature transformation. 
%
%
%
\begin{table}[h]
\caption[ ]{Average stellar parameters for Galactic and LMC RRd stars}
\begin{flushleft}
\begin{tabular}{lrcccc}
\hline
Cluster & N & [Fe/H] & $\log T_{\rm eff}$ & $\log L/L_{\sun}$ 
        & $M/M_{\sun}$ \\
\hline
IC4499 & 13  & $-1.5$ & $\phantom{-}3.830$ & $\phantom{-}1.708$ 
             & $\phantom{-}0.755$ \\
       &     &        & $\pm0.004$ & $\pm0.017$ & $\pm0.021$ \\
M15    &  8  & $-2.3$ & $\phantom{-}3.821$ & $\phantom{-}1.731$ 
             & $\phantom{-}0.730$ \\
       &     &        & $\pm0.004$ & $\pm0.022$ & $\pm0.021$ \\
M68    & 11  & $-2.0$ & $\phantom{-}3.820$ & $\phantom{-}1.728$ 
             & $\phantom{-}0.756$ \\
       &     &        & $\pm0.004$ & $\pm0.013$ & $\pm0.016$ \\
LMC    & 181 & $-1.5$ & $\phantom{-}3.834$ & $\phantom{-}1.736$ 
             & $\phantom{-}0.778$ \\
       &     &        & $\pm0.024$ & $\pm0.132$ & $\pm0.100$ \\
\hline
\end{tabular}
\end{flushleft}
{\footnotesize
\underline {Note:}
Errors listed in the unlabeled rows correspond to the standard deviations 
of the quantities computed for the individual stars (i.e., they are not 
the errors of the averages)}
\end{table}

The first systematic behavior observed from this table is the difference 
in the average temperature between the low- and high-metallicity 
variables. This temperature difference of about 150K is responsible 
mainly for the period difference of $\approx 0\fd05$ between the 
low- and high-metallicity RRd stars. 

By checking the mass and luminosity values, we may suspect some trends  
but they are rather small compared to that of the temperature. This can 
be seen from the very general pulsation relation (e.g., van Albada \& 
Baker 1973)  
%
%
%
\begin{eqnarray}
\log P_0 & = & 11.50 + 0.84\log L/L_{\sun} - 0.68\log M/M_{\sun} \\ \nonumber 
         & - & 3.48\log T_{\rm eff} \hskip 2mm . 
\end{eqnarray}  
Although luminosity and mass differences may contribute to the 
period difference, and, in a more complicated way --- not seen from the 
above simplified equation --- to the period ratios, in the present context 
we assume, as a first approximation, that the masses and luminosities of 
the RRd stars are basically {\it independent} of metallicity. 

Additional support for the assumption of the nearly constant luminosity 
level of the RRd stars can be lent by the narrow range of the observed 
average magnitudes. If we omit the the single outlier 01168, we get 
0.13~mag for the average magnitude range of the remaining 17 RRd stars 
(see Table~1). On the other hand, the same quantity is about 0.4~mag 
for the RRab stars (Kaluzny et al. 1995). 

As far as the mass is concerned, we see that it is only M15 that 
exhibits a somewhat discrepant value. By increasing the metallicity 
of the RRd variables of this cluster to $-2.1$ we get $M/M_{\sun}=0.756$ 
for this cluster, virtually the same value as for the other clusters. 
Although for this metallicity the luminosity increases to $1.747$, 
which is larger than the values of the other clusters, it is still 
within the range of our assumed $\pm 0.02$ ambiguity of the luminosity 
level in the computation of the metallicity (see later). 

At this point we need to refer to the results presented by KW99. 
In that work we did not fix the average cluster metallicities, but 
they were chosen in such a way as to minimize the dispersion between 
the calculated and empirical relative distances (Kov\'acs \& Jurcsik 
1997). This procedure has led to increasing cluster metallicities as 
the metallicity of M15 was increased. Finally, all these resulted in  
systematic differences between the masses of the RRd stars of the 
various clusters. In the interesting metallicity range for M15, we 
obtained $M_{RRd}^{M68}-M_{RRd}^{IC4499}\approx+0.02$ and 
$M_{RRd}^{M15}-M_{RRd}^{IC4499}\approx-0.03$, with some  
sensitivity to changing chemical composition and the assumed 
metallicity of M15. We think that although the optimization 
method of KW99 should be further studied, once even more accurate 
photometric and direct spectroscopic data will be available for 
these and other clusters, a slight difference in the photometric 
zero points could lead to errors in the derived relative cluster 
distances (see Kov\'acs \& Walker 2001), and therefore, to an 
incorrect conclusion about certain intricate details, such as the 
mass differences between the RRd stars.      

The average parameters given in Table~2 for the LMC should be treated 
with caution, because the large period ratio range of its RRd variables 
clearly indicates a substantial spread in [Fe/H]. This is probably the 
main reason why we get a contradictory trend in the luminosity when 
compared with IC4499. Although both systems have the same average 
metallicity, apparently they show a considerable difference in the 
luminosities of their RRd stars. To clarify the situation for the LMC, 
we need a more thorough consideration, which might be difficult with 
the presently available data quality. 

When considering the average masses and luminosities, two additional 
effects might cause some concern. As we have already mentioned in our 
previous papers, the zero point of the temperature scale is the main 
source of the systematic error in the determination of the distance 
modulus with the aid of double-mode variables. We note that the estimated 
ambiguity of $\pm 0.1$~mag in the distance modulus results in $\pm 0.036$ 
and $\pm 0.018$ systematic errors in $\log L$ and $M$, respectively. 
However, these shifts are basically independent of the metallicity, 
and therefore, as we will see next, do not affect our conclusion on 
the metallicity of the Sculptor RRd stars. The other effect on the 
double-mode masses and luminosities comes from a possible change in 
the hydrogen content. By leaving all parameters the same, but changing 
$X$ from $0.76$ to $0.70$, we get systematic changes of $+0.01$ in 
$\log L$ and $+0.024$ in $M$. This effect has a stronger metallicity 
dependence, which yields larger differences in the average RRd masses 
of the various clusters (most of all, because of the relatively low 
adopted metallicity of M15). Nevertheless, alike the temperature 
shift, this composition effect is not essential in the present rough approximation.      

In conclusion, in the following we employ the {\it constant luminosity 
and mass} constraint to the Sculptor RRd variables to derive their 
individual metallicities. For $X=0.76$, we fix the average $\log L$ and 
$M$ values at $1.726$ and $0.755$, respectively (see Table~2). When testing 
the hydrogen dependence, at $X=0.70$ the above quantities are increased 
as mentioned above. In order to take into account possible metallicity 
or other hidden dependence of these averages, we allow $\pm 0.02$ maximum 
errors both in $\log L$ and in $M$.   
%
%
%
   \begin{figure}
   \centering
   \includegraphics[width=90mm]{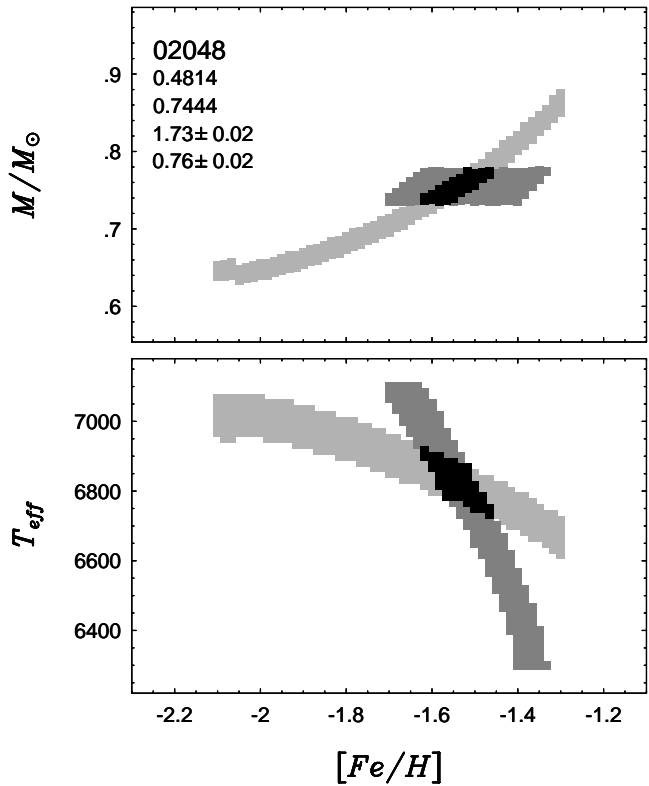}
      \caption{Allowed parameter ranges as functions of the metal 
abundance. Variable name, $P_0$, $P_1/P_0$, average 
$\log L/L_{\sun}$ and $M/M_{\sun}$ values and their ranges are given in 
the left part of the upper panel. Lightest shade corresponds to the case 
when only the luminosity constraint is applied, whereas the darker one 
shows the case when only the mass constraint is used. The darkest shade  
shows the parameters when both constraints are applied. Compositions 
with $X=0.76$ are used} 
         \label{}
   \end{figure}
%
%
%
%
   \begin{figure}
   \centering
   \includegraphics[width=90mm]{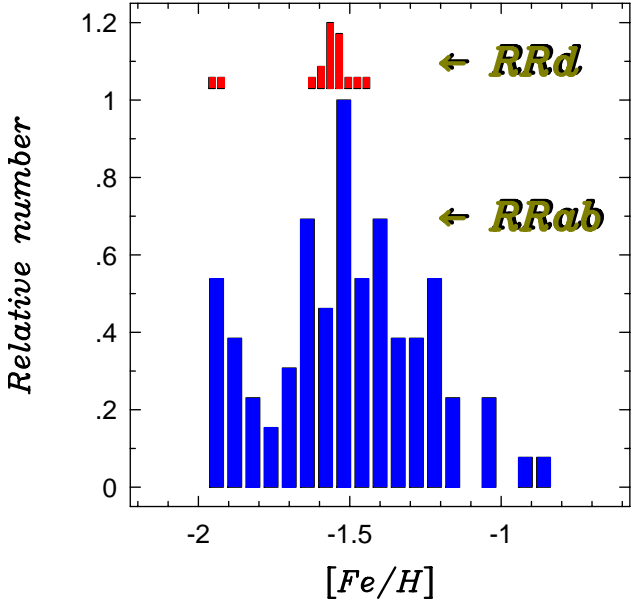}
      \caption{Distribution of [Fe/H] in the Sculptor among the 
RRd and RRab variables. The two distribution functions are normalized 
independently}
         \label{}
   \end{figure}

When the above constraints are applied to the possible parameters obtained 
from the observed periods, we get patterns similar to the one shown in 
Fig.~4. It is seen that none of the constraints, when applied alone, 
is able to give very useful limits on the metallicity, although the 
condition of constant mass yields a much stronger constraint. The very 
different [Fe/H] dependence of the two constraints leads to considerable 
improvement in the accuracy of the metallicity determination when 
both constraints are applied simultaneously. Because of the assumed 
mass and luminosity ranges, the derived metallicities have an ambiguity 
of $\pm 0.07$. (For the low-metallicity stars this error increases by 
a factor of two.) By changing the zero point of the temperature scale 
or using compositions with $X=0.70$, the overlapping regions given by 
the mass and luminosity constraints are shifted vertically by almost 
the same values. This yields the same [Fe/H] values, within $0.02$. 

It is interesting to compare the metallicity distributions of the RRd 
and RRab stars. By using the Fourier decompositions of the light curves 
of the RRab stars (see Jurcsik \& Kov\'acs 1996), Fig.~5 shows the two 
distribution functions. Although some part of the large scatter of the 
RRab metallicities might be accounted for by the large observational 
errors of the light curves, we think that most of the scatter is real, 
and reflects the chemical inhomogeneity of Sculptor. It is noticed that 
the [Fe/H] range of the RRab stars is larger than that estimated from 
the width of the giant branch (e.g., Majewski et al. 1999, hereafter M99) 
and from earlier limited spectroscopic studies (Norris \& Bessel 1978). 
A part of this disagreement could be accounted for by the difference 
of the metallicity scales used in those and in our studies. We use 
the scale defined by high dispersion spectroscopic measurements, which 
is different from the more customary Zinn \& West scale (Zinn \& West 1984). 
If we transform the metallicity range of $[-2.2,-1.4]$ of M99, obtained 
from the width of the giant branch, we get ranges of $[-2.1,-1.2]$ or  
$[-2.3,-1.1]$, depending if we use the transformation formula of 
Carretta \& Gratton (1997, hereafter CG) or Jurcsik (1995, hereafter J95). 
Similarly, by considering the result of M99 derived from evolutionary 
results for the Horizontal Branch stars, for the range of $[-2.3,-1.5]$ 
given by M99, we get ranges of $[-2.3,-1.3]$ (CG) or $[-2.4,-1.3]$ (J95). 
We see that although our estimate for RRab stars still yields larger 
[Fe/H] values by some $\approx 0.2$--$0.3$~dex, the ranges indeed become 
similar when the same metallicity scale is used.

Most of the derived RRd metallicities are close to the average cluster 
value, predicted {\it independently} from the RRab stars. There are two  
RRd variables (01168 and 05354) with considerably lower metallicities 
than the average. We note that these variables are among the securely 
identified RRd stars, and their low metallicities cannot be attributed 
to observational errors. The presence of low metallicity RRd stars 
supports the wide spread of [Fe/H] predicted from the RRab stars. 

Finally we mention two additional results. First, because of the 
two constraints applied, the above computation also yields effective  
temperatures for each variable. Without the two low-metallicity stars, 
for the average effective temperature of the RRd stars we get 
$\approx 6800$K (for the low-metallicity variables we get values lower  
by $\approx 200$K). This can be used as a check of the assumptions made 
in the present derivation, when accurate multicolor photometry will be 
available for this system. 

The second result refers to the possibility that the luminosity and 
mass have a slight dependence on the metallicity. By considering only 
the two representative clusters IC4499 and M68, we may suspect that 
the variables of M68, which have lower metallicities than those of 
IC4499, have larger luminosities (see Table~2). For the mass we may 
keep the assumption of metallicity-independence. If we use these 
constraints, we get some 50K decrease in the temperature, and a very 
small increase of $0.05$ in the metallicity for the Sculptor RRd stars. 
When using chemical compositions with $X=0.70$, the conclusion is very 
similar as far as the changes in the derived metallicity and temperature 
are concerned. It is noted that for this composition we get a metallicity 
effect in the mass, which yields higher masses for higher metallicities 
(the difference is $0.02M_{\sun}$ for the above two clusters).

%
%
 
\section{Conclusions}
The large-scale photometric surveys connected with the search for 
microlensing events in various dense stellar systems, supply valuable 
variable star data both in amount and quality never seen before 
(Paczy\'nsky 2000). In this paper we utilized one of the `side products' 
of these projects. We analyzed the {\sc ogle} database on the Sculptor 
dwarf galaxy (Kaluzny et al. 1995) in order to search for double-mode 
RR~Lyrae (RRd) stars. These variables play an important role both in the 
study of the simplest type of multimode pulsations, and also in the 
estimation of the distances of various stellar systems (Kov\'acs 2000b). 

Frequency analysis of the more than 300 variables in the {\sc ogle} 
database has led to the discovery of 18 RRd stars, from which we 
consider at least 15 as secure identifications. As in the case of 
most of the RRd stars in various systems, all Sculptor variables have 
larger first overtone than fundamental mode amplitudes. Considering 
the periods, these stars are very similar to the ones found in the 
Galactic globular cluster IC4499 (Walker \& Nemec 1996). Together 
with the two long-period stars, they also fit in the overall 
$P_0$---$P_1/P_0$ pattern defined by the RRd stars of the Large 
Magellanic Cloud (Alcock et al. 2000). 

We derived individual metallicities for the Sculptor RRd variables 
by using the results of our former analyses on RRd stars in various 
stellar systems. From these works we obtained luminosity and mass 
values, which, in this first approximation, have been considered to 
be independent of the metallicity. This assumption has led to the 
estimation of the individual metallicities, which span the 
[Fe/H]$\approx[-2.0,-1.5]$ interval, with only two variables at the 
low-metallicity tail with [Fe/H]$\approx -1.9$. In a comparison with 
the independent estimation of the metallicities of the fundamental 
mode RR~Lyrae stars (Jurcsik \& Kov\'acs 1996), it was shown that they 
both lead to the same average metallicity of $-1.5$. It is stressed 
that this conclusion does not depend on the assumed hydrogen content, 
effective temperature zero point and a possible slight dependence of 
the luminosity and mass values on the metallicity. It is also noted 
that our metallicities predicted from fundamental mode RR~Lyrae stars 
are higher by $\approx 0.2$--$0.3$~dex than the estimates given by 
Majewski et al. (1999). On the other hand, when transformed to the 
same metallicity scale, they both yield very similar ranges of 
$\approx 1.0$~dex.  

We recall that the analysis presented in this paper is based on the data 
acquired in the first phase of the {\sc ogle} project which used a 1~m 
telescope for this galaxy with the faintest RRd population known. It is 
clear that multicolor observation of the system with a larger aperture 
telescope would yield a unique opportunity to study the large variable 
star population of this galaxy. With the help of its RRd stars and 
other variables we could estimate its distance, study the metallicity 
distribution on the horizontal branch and investigate the metallicity 
dependence of the luminosity and mass of the RRd stars. It is also 
important to stress the need for direct spectroscopic observations, 
which, except for the first effort in this direction in the Large 
Magellanic Cloud by Clementini et al. (2000), are completely absent.

\begin{acknowledgements}
We are indebted to Janusz Kaluzny and his co-workers for making 
the Sculptor database available to us. Fruitful discussions with 
B\'ela Szeidl, Endre Zsoldos and G\'asp\'ar Bakos are appreciated. 
We thank the referee for his/her constructive comments. The following 
grants are acknowledged: {\sc otka} {\sc t--024022}, {\sc t--026031} 
and {\sc t--030954}. 
\end{acknowledgements}

\end{document}